# A Hybrid Approach to Extract Keyphrases from Medical Documents

Kamal Sarkar*
Computer Science and Engineering Department, Jadavpur University,
Kolkata-700 032, India
jukamal2001@yahoo.com, ksarkar@cse.jdvu.ac.in

## Abstract
Keyphrases are the phrases, consisting of one or more words, representing the important concepts in the articles. Keyphrases are useful for a variety of tasks such as text summarization, automatic indexing, clustering/classification, text mining etc. This paper presents a hybrid approach to keyphrase extraction from medical documents. The keyphrase extraction approach presented in this paper is an amalgamation of two methods: the first one assigns weights to candidate keyphrases based on an effective combination of features such as position, term frequency, inverse document frequency and the second one assign weights to candidate keyphrases using some knowledge about their similarities to the structure and characteristics of keyphrases available in the memory (stored list of keyphrases). An efficient candidate keyphrase identification method as the first component of the proposed keyphrase extraction system has also been introduced in this paper. The experimental results show that the proposed hybrid approach performs better than some state-of-the art keyphrase extraction approaches.

## 1. INTRODUCTION
Medical Literature such as research articles, clinical trial reports, medical case reports, medical news available on the web are the important sources to help clinicians in patient care. The pervasion of huge amount of medical information through WWW has created a growing need for the development of techniques for text summarization, automatic indexing, clustering/classification etc. Keyphrases can be used for such kind of tasks. In a recent work presented in (Wu and Li, 2008)[1], keyphrases have been incorporated in the search results as subject metadata to facilitate information search on the web.

The most keyphrase extraction systems which are proven to be successful have used supervised machine learning techniques. The main advantages of supervised machine learning techniques are that they can adapt to the specific nature of documents at hand based on a representative set of training instances. One of the drawbacks of machine learning based keyphrase extraction systems is that the system is trained on a training set that consists of highly unbalanced distribution of positive (keyphrase) and negative (not a keyphrase) examples because the author provided keyphrases associated with a document are very few in number. The alternative method to overcome this drawback is to label all the phrases in a document. This becomes a difficult task when the dataset is too large and the direct help of domain experts, specifically for the medical domain, is sought.

Considering the above-mentioned problems associated with supervised machine learning based keyphrase extraction techniques, the proposed keyphrase extraction system has been designed based on only the developer/programmers' understanding of the document set and the keyphrase extraction process. This allows the developer/programmers rooms for process tuning based on their understanding of the document set which is relatively small in size compared to the training document-set required for developing a supervised machine learning based techniques. Moreover, the proposed system is not directly affected by the imbalanced distribution of positive and negative examples in the training data. The proposed system uses simple techniques for keyphrase extraction that offers enough insight into the keyphrase extraction process and the factors that affect it.

In this paper, an efficient candidate keyphrase identification method has also been introduced as the first component of the proposed keyphrase extraction system. The candidate keyphrase identification method proposed by us use positional information, domain knowledge and frequency information for choosing the potential candidate keyphrases out of the phrases in a document.

To judge the effectiveness of the proposed method, the proposed system is compared with the state-of-the art keyphrase extraction systems KP-Miner (Elbeltagy and Rafea, 2009)[2] and Kea (Witten, et.al., 1999)[3].

The next section presents a brief description of the previous works related to the work presented in this paper. In section 3, the proposed keyphrase extraction method has been discussed. The evaluation and the experimental results are presented in section 4 and section 5 respectively.

## 2. RELATED WORK
A number of previous works has suggested that document keyphrases can be useful in a various applications such as retrieval engines (Li, et. al., 2004)[4], (Jones and Staveley, 1999)[5], document summarization[6][7], browsing interfaces (Gutwin, et.al., 2003)[8] thesaurus construction (Kosovac, et.al., 2000)[9] and document classification and clustering (Jonse and Mahoui, 2000)[10].

An algorithm to choose noun phrases from a document as keyphrases has been proposed in [11]. Phrase length, its frequency and the frequency of its head noun are the features used in this work. Noun phrases are extracted from a text using a base noun phrase skimmer and an off-the-shelf online dictionary.

Chien [12] developed a PAT-tree-based keyphrases extraction system for Chinese and other oriental languages.

HaCohen-Kerner et al [13][14] proposed a model for keyphrase extraction based on supervised machine learning and combinations of the baseline methods. They applied J48, an improved variant of C4.5 decision tree for feature combination.

Hulth et al [15] proposed a keyphrase extraction algorithm in which a hierarchically organized thesaurus and the frequency analysis were integrated. The inductive logic programming has been used to combine evidences from frequency analysis and thesaurus.

A graph based model for keyphrase extraction has been presented in [16]. A document is represented as a graph in which the nodes represent terms, and the edges represent the co-occurrence of terms. Whether a term is a keyword is determined by measuring its contribution to the graph.

A Neural Network based approach to keyphrase extraction has been presented in [17] that exploits traditional term frequency, inverted document frequency and position (binary) features. The neural network has been trained to classify a candidate phrase as keyphrase or not.

In (Turney, 2000) [18], the problem of keyphrase extraction has been viewed as supervised learning task. In this task, nine features are used to score a candidate phrase; some of the features are positional information of the phrase in the document and whether or not the phrase is a proper noun. Keyphrases are extracted from candidate phrases based on examination of their features. Turney's program is called Extractor. One form of this extractor is called GenEx, which is designed based on a set of parameterised heuristic rules that are fine-tuned using a genetic algorithm. Turney Compares GenEX to the standard machine learning technique called Bagging which uses a bag of decision trees for keyphrase extraction and shows that GenEX performs better than the bagging procedure.

A keyphrase extraction program called Kea, developed by Witten et al. (Witten, et.al, 1999)[3] uses Bayesian learning for keyphrase extraction task. A model is learned from the training documents with exemplar keyphrases and corresponds to a specific corpus containing the training documents. Each model consists of a Naive Bayes classifier and two supporting files containing phrase frequencies and stopped words. The learned model is used to identify the keyphrases from a document. In both Kea and Extractor, the candidate keyphrases are identified by splitting up the input text according to phrase boundaries (numbers, punctuation marks, dashes, and brackets etc.). Finally a phrase is defined as a sequence of one, two, or three words that appear consecutively in the text. The phrases beginning or ending with a stopped word are not taken under consideration. Kea and Extractor both used supervised machine learning based approaches. Two important features: position of the phrase's first appearance into the document and TF*IDF (used in information retrieval setting), are considered for the development of Kea. Here TF corresponds to the frequency of the phrase into the document and IDF is estimated by counting the number of documents in the training corpus that contain the phrase. Witten et al. (Witten, et.al., 1999) [3] compares performance of Kea to Turney's work and shows that performance of Kea is comparable to GenEx proposed by Turney. Moreover, Witten et al. (Witten, et.al., 1999)[3] claims that training Naïve Bayes learning technique is quicker than training GenEx which employs the special purpose genetic algorithm for training. They suggest that deriving domain specific models would be less practical with the original lengthy genetic algorithm approach.

A effective keyphrase extraction system called KP-Miner developed by El-Beltagy et.al.(Elbeltagy and Rafea, 2009)[2] uses simple features like term (phrase) frequency (TF), inverse document frequency(IDF) and position of a phrase(that is, whether a phrase appears early in the document or not). One important feature used in this keyphrase extraction system is the boosting factor which is used to boost up the TF*IDF weight of the multi-word phrases. This is based on the fact that frequency of a multi-word phrase in a small corpus is less than that of a single word phrase. This issue has also been addressed in (Sarkar, 2011) [19] in slightly different way.

A neural network based keyphrase extraction system has been presented in (Sarkar, el. al., 2010) [20]. This uses features like phrase frequency, phrase links to other phrases, inverse document frequency, phrase position, phrase length and word length.

Keyphrase extraction using Naïve Bayes in medical domain has been presented in (Sarkar, 2009)[21]. This system has been tested on a small set of 25 documents.

The work presented in (Li and Brook Wu, 2006) [22] uses MeSH (Medical Subject Headings) as a knowledge base to determine domain specificity of a phrase.

## 3. PROPOSED APPROACH TO KEYPHRASE EXTRACTION

The proposed keyphrase extraction method consists of three primary components: document pre-processing, candidate keyphrase identification and assigning scores to the candidates for ranking.

### 3.1. Document preprocessing
The pre-processing task includes formatting the document. If the source document is in pdf format, it is converted to text format before submission to the keyphrase extractor.

### 3.2. Candidate keyphrase identification
A simple and knowledge poor approach to candidate keyphrase identification is adopted as the first step of the proposed system. This approach is a variant of the candidate keyphrase identification approach presented in (Kumar and Srinathan, 2008) [23]. A candidate keyphrase is considered as a sequence of words containing no punctuations and stop words. A list of common verbs is also added to the stop word list because it is observed that the author assigned keyphrases rarely contains common verbs. The process of candidate keyphrase extraction has two steps:

Step1: extraction of candidate keyphrases considering punctuations and stop words as the phrase boundary,

Step2: Breaking further the phrases selected at the step one into smaller phrases using the following rules:

i. If a phrase is L-word long, all n-grams (n varies from 1 to L-1) are generated and added to the candidate phrase list.

ii. If a phrase is longer than 5 words, it is discarded.

Figure 1 shows a sample sentence and the candidate keyphrases identified from this sentence. Some candidate phrases generated using the above mentioned method may not be meaningful to human readers. For example, in figure1, the candidate phrase "investigate potential" is less meaningful. After computing phrase frequency and phrase weight, such kind of candidate keyphrases are filtered out. For this purpose, two conditions are applied. Condition one is to choose threshold on the phrase weight (Phrase weighting scheme has been presented in the next subsection) which is a function of phrase frequency, inverse document frequency, domain knowledge etc.

The second condition is related to the first appearance of the phrase in the document. Previous works (Witten, et.al., 1999) [3] have suggested that keyphrases appear sooner in an article. The works in (Elbeltagy and Rafea, 2009) [2] states that a phrase occurring the first time after a predefined threshold is less likely a keyphrase. A threshold is set on the position of the phrases where the phrases are numbered sequentially and the first phrase in the document is numbered as 1 and the last phrase is numbered as N. If a phrase appears first after the given threshold it is ignored, that is, if a phrase X appears first at pos i and the threshold value is set to $T_{pos}$ and $i > T_{pos}$, the phrase X is discarded.

---

**Sample Sentence:**

*This study was one of the first to investigate potential risk factors for anxiety (i.e., behavioral inhibition, parental negative affect, parenting stress) in early childhood.*

**Initial list of candidate keyphrases (after step1).**

study, investigate potential risk factors, anxiety, behavioral inhibition, parental negative affect, parenting stress, childhood

**The list of candidate phrases (after step2).**

study, investigate, potential, risk, factors, investigate potential, potential risk,  risk factors, investigate potential risk, potential risk factors, anxiety, behavioral inhibition, behavioral, inhibition, parental negative affect, parental, negative, affect, parental negative, negative affect, parenting stress, parenting, stress, childhood

---

**Fig 1: A sample sentence and the candidate keyphrases identified from this sentence**

## 3.3. Assigning scores to candidate keyphrases

In general, a document has a few number of author assigned keyphrases. To select a small subset of candidates as the keyphrases requires assigning weights to the candidates and raking them based on these weights.

The weight of a candidate keyphrase is computed using three important features: phrase frequency, inverse document frequency and domain specificity.

### 3.3.1. Weighting using phrase frequency (PF) and inverse document frequency (IDF)

The score for a candidate keyphrase due to PF and IDF features is computed using the following formula:

$$SCORE_{pf*idf} = \begin{cases} PF*IDF, & \text{if plength=1} \\ PF*\log(N), & \text{if plength>1} \end{cases} \quad (1)$$

Where:

plength= length of the phrases in terms of words

PF = phrase frequency which is counted as number of times a phrase occurs in a document

IDF= log(N/DF), where N is the total number of documents in the corpus(a collection of documents in a domain under consideration) and DF is the number of documents in which a phrase occurs at least once. Equation (1) shows that for multi-word phrases, phrase score is computed using PF * log (N), which is basically PF * IDF with DF set to 1. This is due to the fact that multi-word phrases do not occur as frequently within a relatively small collection of documents as do single-word phrases.

### 3.3.2. Using domain knowledge for weighting candidate keyphrases

A score is assigned to a candidate keyphrase based on how much it is similar to the structure and characteristics of keyphrases available in the memory (a stored list of keyphrases for the domain under consideration). For this purpose, a keyphrase list is created with readily available author assigned keyphrases collected from medical journal articles. A keyphrase list of 1940 Keyphrases is used to create a domain specific glossary database giving some knowledge about the structure and characteristics of keyphrases. The documents wherefrom this list of keyphrases are collected for creating glossary database are not included in the set of documents (the test set) on which the proposed keyphrase extraction system is tested. However, using such a list of keyphrases stored in the memory for weighing the candidate keyphrases can be considered as some sort of partial supervision provided to the keyphrase extraction system. The use of this kind of knowledge base in keyphrase extraction task has previously been investigated in [25] [1]. A variant of the method presented in [1] is used for the proposed domain specific keyphrase extraction task.

From the keyphrase list, two tables are created: table1 is the keyword table which is created by splitting the keyphrases belonging to the keyphrase list into words that can be called as keywords. This table has two columns (keyword, weights) and table2 is key sub-phrase table which consists of all sub-phrases generated from the keyphrases in the keyphrase list. For any manual keyphrase in the keyphrase list, all possible n-grams (n varies from 2 to n) are generated and included in key sub-phrase table. The key sub-phrase table has also two columns (sub-phrase, weights).

Weights for keywords in the keyword table are assigned using the following rules:

- If a keyword appears always alone independently in the keyphrase list it is assigned a score of 1.

- If the keyword appears always as part of another keyphrase, that is, if it has no independent existence in the keyphrase list, it is assigned a score which is computed as 1/log(c), where c is the number of times the keyword appears as the part of keyphrases. Here it is assumed that a keyword, which has no independent existence and repeats many times in the keyphrase-list only as the parts of other keyphrases, is less domain-specific.

- If the keyword appears independently in the keyphrase list in some cases and also appears as part of keyphrases in some other cases, it is assigned a score which is computed based on the formula: 0.5 * ( 1 + 1 / log (c) ), where c is the number of times the keyword occurs as the part of keyphrases.

Weight of a sub-phrase or a phrase in the key sub-phrase table is computed by summing up weights of keywords of the sub-phrase or the phrase.

The keyword table and the key sub-phrase table are used as domain knowledge in computing a score for a candidate keyphrase. The score for a candidate phrase is computed using the following equation:

$$SCORE_D = \sum_{i=1}^{M} K_i + \sum_{j=1}^{PC} P_j \quad (2)$$

Where:

$K_i$ = the weight of the i-th keyword in the candidate keyphrase.

$P_j$ = weight of the j-th sub-phrase associated with the candidate keyphrase

M = the number of keywords in a candidate keyphrase

PC = the number of sub-phrases generated from the candidate keyphrase

The sub-phrases of the candidate keyphrase are generated by computing all possible n-grams, where n varies from 2 to length of the phrase. When n is set to the length of the candidate keyphrase, the n-gram is basically the candidate keyphrase. The reason for taking the weights of all possible sub-phrases in calculating the candidate keyphrase score, in addition to the weights of individual words, is to decide whether a sub-phrase is a manual keyphrase in the keyphrase table. If it is, this candidate keyphrase is assumed to be more important. This feature will favor those candidate keyphrases which itself or whose parts are found in the knowledge base. Availability of a phrase or its sub-phrases in the knowledge base provides some evidence in support of keyphrase worthiness of a phrase. Thus, with this knowledge base, the proposed keyphrase extraction system is provided with some sort of partial supervision.

*3.3.3. Combining two weighting schemes*
Two weighting schemes have already been discussed in the previous subsections. These two types of scores should be combined to assign a unique score to each candidate keyphrase. The combined score for a candidate keyphrase is computed using the following linear combination of two scores.

$$SCORE = \alpha * SCORE_{pf*idf} + (1-\alpha) * SCORE_D \quad (3)$$

Where:

$SCORE_{pf*idf}$ is the score based on phrase frequency and inverse document frequency computed using the equation (1).

$SCORE_D$ is the score based on domain specificity of a phrase computed using the equation (2).

$\alpha$ is the tuning parameter whose value is decided through experimentation. The best results are obtained when $\alpha$ is set to 0.6.

### 3.4. Extracting keyphrases
After assigning scores to the candidate keyphrases, the next step is to select K top-ranked candidate keyphrases as the final list of keyphrases. The value of K is specified by the user.

## 4. EVALUATION

For the evaluation of the performance of the proposed system, keyphrases extracted by the proposed system are compared with those assigned by the author. The three evaluation metrics are used: Precision, recall and the average number of key- phrases extracted correctly per document (Average Keys). The keyphrases assigned by the original author(s) are used as the standard keyphrase set. The system-generated keyphrases are compared to the author assigned keyphrases. Precision is defined as the proportion of the extracted keyphrases that match the keyphrases assigned by a document's author(s). Recall is defined as the proportion of the keyphrases assigned by a document's author(s) that appear in the set of keyphrases generated by the keyphrase extraction system.

For system evaluation when a system generated keyphrase is compared with author assigned keyphrases, a match is considered to have been found if the stem of the system-generated keyphrase matches the stem of any of the author-assigned keyphrases where stemming is carried out via porters algorithm (Porter, 1980)[24].

Measuring precision and recall against author provided keyphrases allows comparisons between different keyphrase extraction systems.

To test the proposed keyphrase extraction system, 300 medical journal articles have been downloaded from a number of online medical journals such as Indian Journal of Surgery, Journal of General Internal Medicine, The American Journal of Medicine, International Journal of Cardiology, Journal of Anxiety Disorder. The downloaded research articles are basically available as PDF files. All PDF files are converted to text files. Author assigned keywords associated with the test documents are separated from the documents while the documents are submitted to the keyphrase extraction system. The documents on which the proposed system is tested are not considered while creating the glossary database as discussed in subsection 3.3.2.

The average number of author-assigned keyphrases over all articles in the test corpus used for the proposed work is 4.331.

## 5. EXPERIMENTS AND RESULTS

For extracting the keyphrases from a test document, the proposed system identifies first the candidate keyphrases,

computes phrase weight using the equation(3), filters out noisy phrases based on two conditions discussed in subsection 3.2 and assigns scores to the remaining candidate keyphrases. Finally the top-ranked K candidate keyphrases are selected as keyphrases. To filter out noisy phrases, two conditions discussed in subsection 3.2 are applied here in the pre-specified order as follows:

(1) Phrases whose position of the first occurrence in the document is greater than $T_{pos}$ are discarded. The value of $T_{pos}$ is set to 120 to obtain the best results on the dataset used for the proposed work, (2) the threshold value on the phrase weight is adjusted to keep those candidate keyphrases which occurs at least twice in a document or which has higher similarity to the phrases in the manually created knowledge base.

**Table 1 System performance comparisons based on average number of extracted keyphrases that match with author assigned Keyphrases (SD means standard deviation)**

| # of KEYS | Proposed method | KP-Miner | Kea |
|---|---|---|---|
| | *Average Keys ± SD* | *Average Keys ± SD* | *Average Keys ± SD* |
| 5 | **1.479 ± 0.967** | 1.273 ± 0.940 | 1.050 ± 0.9385 |
| 10 | **1.810 ± 1.164** | 1.70 ± 1.152 | 1.661 ± 1.115 |
| 15 | **2.100 ± 1.246** | 1.959 ± 1.158 | 1.884 ± 1.1774 |

To prove the effectiveness of the proposed keyphrase extraction method, the proposed system is compared to 2 top performing keyphrase extraction systems, KP-Miner (Elbeltagy and Rafea, 2009)[2] and Kea (Witten, et.al., 1999)[3].

El-Beltagy et.al (Elbeltagy and Rafea, 2009)[2] have shown that the system, KP-Miner performs better than two publicly available systems Kea[3] and Extractor (Turney, 2000) [18].

The version 5.0 of Kea[1] is downloaded and installed on a machine for testing. Kea is trained with 50 medical documents (the amount of trained documents was recommended by Kea's developers [22]) and the associated keyphrases. The data set used for training Kea is independent of the test set of 300 documents. After training Kea, a model is built based on Naïve Bayes. This pre-built model is used to extract keyphrases from the test documents.

To experiment with KP-Miner, the demo version of this system available at the website http://www.claes.sci.eg/coe_wm/kpminer/ is used.

---

[1] http://www.nzdl.org/Kea/

Two existing systems KP-Miner and Kea and the proposed method have been tested on the test set of 300 medical documents.

The precision and recall for all the above-mentioned systems are calculated when the number of extracted keyphrases is 5, 10 and 15 respectively.

The table 1 shows that the proposed keyphrase extraction method outperforms two existing keyphrase extraction systems: KEA and KP-Miner. Table 2 compares, in terms of precision and recall, the performances of the proposed keyphrase extraction system, the system called KP Miner [2] and the system called Kea [3]. The results shown in table 2 and table 1 indicate that the proposed keyphrase extraction method performs better than other two systems to which it is compared.

**Table 2 Comparisons of system performances based on Precision (Pre) and Recall (Re) (SD means standard deviation)**

| # of kEYS | Proposed Method | | KP Miner | | Kea | |
|---|---|---|---|---|---|---|
| | Pre ± SD | Re ± SD | Pre ± SD | Re ± SD | Pre ± SD | Re ± SD |
| 5 | **0.296 ± 0.185** | **0.352 ± .228** | 0.255 ± 0.188 | 0.310 ± 0.236 | 0.210 ± 0.188 | 0.248 ± 0.218 |
| 10 | **0.181 ± 0.116** | **0.434 ± 0.272** | 0.170 ± 0.115 | 0.408 ± 0.275 | 0.166 ± 0.111 | 0.395 ± 0.249 |
| 15 | **0.138 ± 0.083** | **0.49 ± 0.278** | 0.131 ± 0.077 | 0.465 ± 0.277 | 0.126 ± 0.079 | 0.448 ± 0.271 |

## 6. CONCLUSION

This paper discusses a hybrid keyphrase extraction approach in medical domain. The proposed approach combines domain knowledge with the features namely phrase frequency, inverse document frequency and phrase position in more effective way. An efficient candidate keyphrase identification component has also been used as the first part of the proposed keyphrase extraction system. The proposed approach results in an easy-to-implement keyphrase extraction system that outperforms some state-of-the art keyphrase extraction systems. The experimental results also suggest that the proposed keyphrase extraction method is effective in medical domain and incorporation of domain knowledge as partial supervision boosts up the system performance.